\documentclass[prd,superscriptaddress,amsfonts,amssymb,amsmath,showpacs,twocolumn]{revtex4-1}
\usepackage{bm}
\usepackage{amsfonts}
\usepackage{latexsym}
\usepackage[latin1]{inputenc}
\usepackage{graphicx}
\usepackage{amsmath}
\usepackage{palatino}
\usepackage{mathpazo}
\usepackage{textcomp}
\usepackage{float}
\usepackage{booktabs}
\usepackage{dcolumn}
\usepackage{hyperref}
\hypersetup{colorlinks,citecolor=blue}
\usepackage{amsmath}
\usepackage{xcolor}
\usepackage{orcidlink}
\usepackage[caption=false]{subfig}
\usepackage{commath}
\def\jnl@style{\it}
\def\aaref@jnl#1{{\jnl@style#1}}

\def\aaref@jnl#1{{\jnl@style#1}}

\def\aj{\aaref@jnl{AJ}}                   
\def\apj{\aaref@jnl{ApJ}}                 
\def\apjl{\aaref@jnl{ApJ}}                
\def\apjs{\aaref@jnl{ApJS}}               
\def\apss{\aaref@jnl{Ap\&SS}}             
\def\aap{\aaref@jnl{A\&A}}                
\def\aapr{\aaref@jnl{A\&A~Rev.}}          
\def\aaps{\aaref@jnl{A\&AS}}              
\def\mnras{\aaref@jnl{Mon.~Not.~Roy.~Astron.~Soc.}}             
\def\prd{\aaref@jnl{Phys.~Rev.~D}}        
\def\prc{\aaref@jnl{Phys.~Rev.~C}}  
\def\prl{\aaref@jnl{Phys.~Rev.~Lett.}}    
\def\qjras{\aaref@jnl{QJRAS}}             
\def\skytel{\aaref@jnl{S\&T}}             
\def\ssr{\aaref@jnl{Space~Sci.~Rev.}}     
\def\zap{\aaref@jnl{ZAp}}                 
\def\nat{\aaref@jnl{Nature}}              
\def\aplett{\aaref@jnl{Astrophys.~Lett.}} 
\def\apspr{\aaref@jnl{Astrophys.~Space~Phys.~Res.}} 
\def\physrep{\aaref@jnl{Phys.~Rep.}}      
\def\physscr{\aaref@jnl{Phys.~Scr}}       
\def\commat{\aaref@jnl{Comm.~Math.~Phys.}}              
\def\science{\aaref@jnl{Science}}               
\def\cqg{\aaref@jnl{Classical Quant.~Grav.}}            
\def\jpcs{\aaref@jnl{JPCS}}                                     
\def\ijmpd{\aaref@jnl{Int.~J.~Mod.~Phys.~D}}                    
\def\grg{\aaref@jnl{Gen.~Relat.~Gravit.}}               
\def\rpp{\aaref@jnl{Rep.~Prog.~Phys.}}          
\def\npa{\aaref@jnl{Nucl.~Phys.~A}}        
\def\lrr{\aaref@jnl{Living Rev.~Rel.}}                   
\def\jcap{\aaref@jnl{J.~Cosmology Astropart.~Phys.}}    
\def\rmp{\aaref@jnl{Rev.~Mod.~Phys.}}   
\def\epjc{\aaref@jnl{Eur.~Phys.~J.~C}}

\allowdisplaybreaks[1]
\renewcommand{\arraystretch}{1.1}
\begin{document}
\color{red}

\title{Cosmography in $f(Q)$ gravity}

\author{Sanjay Mandal\orcidlink{0000-0003-2570-2335}}
\email{sanjaymandal960@gmail.com}
\affiliation{Department of Mathematics, Birla Institute of Technology and
Science-Pilani,\\ Hyderabad Campus, Hyderabad-500078, India.}
\author{Deng Wang\orcidlink{0000-0003-2062-5828}}
\email{cstar@nao.cas.cn}
\affiliation{National Astronomical Observatories, Chinese Academy of Sciences, Beijing, 100012, China}
\author{P.K. Sahoo\orcidlink{0000-0003-2130-8832}}
\email{pksahoo@hyderabad.bits-pilani.ac.in}
\affiliation{Department of Mathematics, Birla Institute of Technology and
Science-Pilani,\\ Hyderabad Campus, Hyderabad-500078, India.}

\date{\today}
\begin{abstract}
Cosmography is an ideal tool to investigate the cosmic expansion history of the Universe in a model-independent way. The equations of motion in modified theories of gravity are usually very complicated; cosmography may select practical models without imposing arbitrary choices a priori. We use the model-independent way to derive $f(z)$ and its derivatives up to fourth order in terms of measurable cosmographic parameters. We then fit those functions into the luminosity distance directly. We perform the MCMC analysis by considering three different sets of cosmographic functions. Using the largest supernovae Ia Pantheon sample, we derive the constraints on the Hubble constant $H_0$ and the cosmographic functions, and find that the former two terms in Taylor expansion of luminosity distance work dominantly in $f(Q)$ gravity. 
\end{abstract}

\maketitle

\section{Introduction}\label{sec1}
At the beginning of the 20th century, Albert Einstein proposed the General Theory of Relativity (GR), which changes our understanding of the Universe. It is growing by a prominent number of correct observations and exploring the hidden scenarios of the Universe in modern cosmology. Later on, a group of supernovae observations confirms that our Universe is currently going through the accelerated expansion phase \cite{riess/1998}. This causes the existence of high negative pressure in the Universe, which produces by the unknown form of energy and matter called dark energy (DE) and dark matter (DM). To know the unknown form of energy is a challenging task for the researchers in the modern era. In GR, the cosmological constant, $\Lambda$ is the simplest candidate, which explains the vacuum energy \cite{carroll/1992}. Rather than this, it fails to overcome some problems such as age problem \cite{yang/2010}, coincidence problem \cite{veltan/2014}. In this regard, it is considered that GR may not be the correct proposal to describe gravity in large-scale structures. Modified gravity theories are the generalization of the general theory of relativity, and it violates the Strong Equivalence Principle (SEP) \cite{joyce/2016}. Despite little progress so far in understanding cosmic acceleration \cite{baker/2019}, modified gravity studies are important as they provide reliable, logical alternatives to GR and may ease some of the current problems. In the last two decades, many works are carried out in modified gravity theories to explore and understand the profile of the Universe (see the references \cite{peracaula/2019}).

In GR, we use the Livi-Civita connection to describe its gravitational interaction in Riemannian space-time. This choice is built on the hypothesis of free geometry of torsion and nonmetricity. Besides this, we have to keep in mind that the general affine connection has a more generic expression \cite{hehl/1995}, and GR can also be derived in different space-time other than Riemannian. Teleparallel gravity is an alternative theory to GR, whose gravitational interaction is described by the torsion, $T$ \cite{aldrovandi/2014}.  Its Teleparallel Equivalent of General Relativity (TEGR) used the Weitzenbock connection, which implies zero curvature and nonmetricity \cite{maluf/2013}. The Weitzenbock connection of the disappearance of the sum of the curvature and the scalar torsion is considered in Weyl-Carten space-time for a cosmological model \cite{haghani/2012}. The motion equations can be derived from the Einstein-Hilbert type of variational principle, and they completely depend on the Lagrange multiplier (see details \cite{haghani/2013}). The case of Riemann-Cartan space-times with zero nonmetricity, which mimics the teleparallel theory of gravity, was also considered. Symmetric teleparallel gravity is also an alternative theory, where zero curvature and torsion are considered \cite{nester/1999}. In this theory, the nonmerticity, $Q$ described the gravitational interaction. In the last decades, researchers are attracted towards the modified theories of gravity, because it describes the current phenomenon of the Universe. As a result, the gravitational interactions have been derived by using different types of geometrics \cite{ferraro/2008}.

Moreover, by assuming the affine connection, one specifies the metric-affine geometry \cite{jarv/2018}. As we know, the metric tensor $g_{\mu\nu}$ is the generalization of gravitational potential, used to define the distances, angles, and volumes, whereas the affine connection defines the covariant derivatives and parallel transport. In differential geometry, the general affine connection can be written as
\begin{equation}\label{1a}
{\Gamma^{\lambda}}_{\mu\nu}=\left\lbrace{^{\lambda}}_{\mu\nu}\right\rbrace+{K^{\lambda}}_{\mu\nu}+{L^{\lambda}}_{\mu\nu},
\end{equation}
where the Levi-Civita connection of the metric is
\begin{equation}\label{1b}
\left\lbrace{^{\lambda}}_{\mu\nu}\right\rbrace\equiv\frac{1}{2}g^{\lambda \beta}\left(\partial_{\mu}g_{\beta \nu}+\partial_{\nu}g_{\beta \mu}-\partial_{\beta}g_{\mu \nu}\right),
\end{equation}
the contortion is
\begin{equation}\label{1c}
{K^{\lambda}}_{\mu\nu}\equiv\frac{1}{2}{T^{\lambda}}_{\mu\nu}+T_{(\mu}{}^{\lambda}{}_{\nu)},
\end{equation}
with the torsion tensor ${T^{\lambda}}_{\mu\nu}\equiv 2{\Gamma^{\lambda}}_{[\mu\nu]}$, and the disformation $L^{\lambda}_{\mu\nu}$ can be written in terms of the nonmetricity tensor as
\begin{equation}\label{1d}
{L^{\lambda}}_{\mu\nu}\equiv \frac{1}{2} g^{\lambda\beta}\left(-Q_{\mu\beta\nu}-Q_{\nu \beta\mu}+Q_{\beta\mu\nu}\right).
\end{equation}
Here, the non-metricity tensor $Q_{\gamma \mu\nu}$ is defined as the minus covariant derivative of the metric tensor with respect to the Weyl-Cartan connection ${\Gamma^{\lambda}}_{\mu\nu}$, $Q_{\gamma \mu\nu}\equiv \nabla_{\gamma}g_{\mu\nu}$, and it can be written as \cite{hehl/1976};
\begin{equation}\label{1e}
Q_{\gamma \mu\nu}=-\frac{\partial g_{\mu\nu}}{\partial x^{\gamma}}+g_{\nu \lambda}{\Gamma^{\lambda}}_{\mu\gamma}+g_{\lambda\mu}{\Gamma^{\lambda}}_{\nu\gamma}.
\end{equation}

In this article, we will work on symmetric teleparallel gravity in which the gravitational interaction is completely described by the nonmetricity $Q$ with torsion and curvature free geometry. As this is a novel approach to exploring some Universe insights, so far, a few works have been done in this approach. Exploring this formulation will hopefully provide some insight into the current scenario of the Universe. R. Lazkoz et al., have been analyzed the different form of $f(Q)$ by transferring it to redshift form $f(z)$ with observational data. They proposed various polynomial forms of $f(z)$ including additional terms which causes the deviation from $\Lambda$CDM model and checks their validity \cite{Lazkoz/2019}. S. Mandal et al., studied the energy conditions in order to check the stability of their assumed cosmological models and constraints the model parameters with the present values of cosmological parameters in $f(Q)$ gravity \cite{mandal/2020}. Jianbo Lu et al., studied symmetric teleparallel gravity comparing with $f(T)$ and $f(R)$ gravity, and found some interesting results. Besides, they investigated one $f(Q)$ model and shown five critical points in the STG model \cite{jianbo/2019}. Mikhel R\"unkla and Ott Vilson \cite{mikhel/2018}, V. Gakis et al. \cite{viktor/2020}, studied the extension of symmetric teleparallel gravity in which they have been reformulated the scalar non-metricity theories, derived the field equations, and discussed their properties. T. Harko et al. \cite{harko/2018}, in their interesting work, they have been proposed the extension of symmetric teleparallel gravity by considering the Lagrangian of the form of non-minimal coupling between the non-metricity $Q$ and the matter Lagrangian. Besides this, they have been studied several cosmological aspects by presuming power law and exponential forms of $f_1(Q)$ and $f_2(Q)$. They also found that their model shows the accelerated expansion of the universe. The motivation of working in symmetric teleparallel gravity is that in this approach, the field equations are in second-order, which is easy to solve. Furthermore, the advantage is that it overcomes the problem which is generated by the higher derivative property of the scalar $R$ such as for a density of a canonical scalar field $\phi$, the non-minimal coupling between geometry and matter Lagrangian produces an additional kinetic term which is not an agreement with the stable Horndeski class \cite{olmo/2015}. In this work, we focus on constraint the functions of the cosmographic set using the cosmographic idea, which provides the maximum amount of information from the luminosity distances of SNe Ia. To constrain those functions, we have adopted a Bayesian statistical analysis using MCMC simulation with the latest large Pantheon data set. 

The outlines of the paper are as follows. In Sec. \ref{sec2}, we discuss the Einstein Lagrangian for the symmetric teleparallel geometry. We briefly present the formalism of $f(Q)$ gravity in Sec. \ref{sec3}. In Sec. \ref{sec4}, we have discussed the cosmographic parameters with its origin and use. After this, we have expressed the $f(Q)$ and its derivatives in terms of cosmographic parameters in Sec. \ref{sec5}. Then, we have constraints three models of function of cosmographic set in Sec. \ref{sec6}. There we use MCMC method to constraint the parameters with the latest Pantheon data set. Finally, we conclude our outcomes in Sec. \ref{sec7}.

\section{Covariant Einstein Lagrangian}\label{sec2}

Albert Einstein presented a simple Lagrangian for his motion equations using the Levi-Civita connection defined in Eq. \ref{1b}, in 1916 \cite{einstein/1916}, which is given by
\begin{equation}\label{2a}
L_E=g^{\mu\nu}\left(\left\lbrace{^{\alpha}}_{\beta\mu}\right\rbrace \left\lbrace{^{\beta}}_{\nu\alpha}\right\rbrace - \left\lbrace{^{\alpha}}_{\beta\alpha}\right\rbrace \left\lbrace{^{\beta}}_{\mu\nu}\right\rbrace\right)
\end{equation}
Nevertheless, the standard Lagrangian formulation was proposed by Hilbert in 1915. The Lagrangian is described by the Ricci scalar $\mathcal{R}$, which contains the metric tensor's second-order derivatives. Moreover, the Ricci scalar for this formulation can be written as
\begin{equation}\label{2b}
\mathcal{R}=L_E+L_B,
\end{equation}
where $L_B$ is the boundary term, and it is given by
\begin{equation}\label{2c}
L_B=g^{\alpha\mu}\mathcal{D}_{\alpha}\left\lbrace{^{\nu}}_{\mu\nu}\right\rbrace-g^{\mu\nu}\mathcal{D}_{\alpha}\left\lbrace{^{\alpha}}_{\mu\nu}\right\rbrace
\end{equation}
The symbol $\mathcal{D}_\alpha$ represents the covariant derivative with the Levi-Civita connection \ref{1b}. The Lagrangian defined in Eq. \ref{2a} is not a covariant one; therefore, the higher-order derivative mechanism result the standard one. Also, one can upgrade the Christoffel symbol to a covariant one using partial derivatives. Hence, one can write Eq. \ref{1d} with the covariant derivative $\nabla_\alpha$ as
\begin{equation}\label{2d}
{L^{\alpha}}_{\beta\gamma}=-\frac{1}{2}g^{\alpha\lambda}\left(\nabla_{\gamma}g_{\beta\gamma}+\nabla_{\beta}g_{\lambda\gamma}-\nabla_{\lambda}g_{\beta\gamma} \right).
\end{equation}
Now, the non-metricity, $Q$, can be written as
\begin{equation}\label{2e}
Q=-g^{\mu\nu}\left({L^{\alpha}}_{\beta\mu}{L^{\beta}}_{\nu\alpha}-{L^{\alpha}}_{\beta\alpha}{L^{\beta}}_{\mu\nu}\right).
\end{equation}
Whenever the covariant derivative reduces to the partial derivative at that time, the non-metricity $Q$ will be equivalent to the negative of the Einstein Lagrangian \ref{2a} i.e.
\begin{equation}
\nabla_{\alpha}\circeq {\partial} _{\alpha}, \hspace{0.5cm} Q\circeq -L_{E},
\end{equation}
where $`o'$ in the above expressions was called the \textit{gauge coincident}, it is consistent in the symmetric teleparallel geometry \cite{Jimenez/2018}.\newline
In symmetric teleparallel geometry, the connection ${\Gamma^{\alpha}}_{\mu\nu}$ does not depend on the curvature and torsion. However, the connection in Eq. \ref{1b} and its curvature still show their physical roles. Remember that the Dirac Lagrangian, connected with the connection ${\Gamma^{\alpha}}_{\mu\nu}$ in the symmetric teleparallel geometry, filters out everything but the Christoffel symbols \ref{1b} from ${\Gamma^{\alpha}}_{\mu\nu}=\left\lbrace{^{\alpha}}_{\mu\nu}\right\rbrace+{L^{\alpha}}_{\mu\nu}$. As a consequence, the symmetric teleparallel mechanism is a good and stable modification of GR. Since (minimally coupled) fermions are still metrically connected  \cite{tomi/2018}, and although the pure gravity field is now trivially interconnected, nothing actually changes, but only the higher-derivative boundary term $L_B$ disappears from this operation.

\section{Motion Equations in $f(Q)$ gravity}\label{sec3}

Let us consider the action for $f(Q)$ gravity given by \cite{Jimenez/2018}
\begin{equation}
\label{1}
\mathcal{S}=\int \frac{1}{2}\,f(Q)\,\sqrt{-g}\,d^4x+\int \mathcal{L}_m\,\sqrt{-g}\,d^4x\,,
\end{equation}
where $f(Q)$ is a general function of the $Q$, $\mathcal{L}_m$ is the matter Lagrangian density and $g$ is the determinant of the metric $g_{\mu\nu}$.\\
The nonmetricity tensor and its traces are such that
\begin{equation}
\label{2}
Q_{\gamma\mu\nu}=\nabla_{\gamma}g_{\mu\nu}\,,
\end{equation}
\begin{equation}
\label{3}
Q_{\gamma}={{Q_{\gamma}}^{\mu}}_{\mu}\,, \qquad \widetilde{Q}_{\gamma}={Q^{\mu}}_{\gamma\mu}\,.
\end{equation}
Moreover, the superpotential as a function of nonmetricity tensor is given by
\begin{equation}
\label{4}
4{P^{\gamma}}_{\mu\nu}=-{Q^{\gamma}}_{\mu\nu}+2Q_{({\mu^{^{\gamma}}}{\nu})}-Q^{\gamma}g_{\mu\nu}-\widetilde{Q}^{\gamma}g_{\mu\nu}-\delta^{\gamma}_{{(\gamma^{^{Q}}}\nu)}\,,
\end{equation}
where the trace of nonmetricity tensor \cite{Jimenez/2018} has the form
\begin{equation}
\label{5}
Q=-Q_{\gamma\mu\nu}P^{\gamma\mu\nu}\,.
\end{equation}
Another relevant ingredient for our approach is the energy-momentum tensor for the matter, whose definition is
\begin{equation}
\label{6}
T_{\mu\nu}=-\frac{2}{\sqrt{-g}}\frac{\delta(\sqrt{-g}\mathcal{L}_m)}{\delta g^{\mu\nu}}\,.
\end{equation}
Taking the variation of action \eqref{1} with respect to metric tensor, one can find the field equations
\begin{widetext}
\begin{equation}
\label{7}
\frac{2}{\sqrt{-g}}\nabla_{\gamma}\left( \sqrt{-g}f_Q {P^{\gamma}}_{\mu\nu}\right)+\frac{1}{2}g_{\mu\nu}f
+f_Q\left(P_{\mu\gamma i}{Q_{\nu}}^{\gamma i}-2Q_{\gamma i \mu}{P^{\gamma i}}_{\nu} \right)=-T_{\mu\nu}\,,
\end{equation}
\end{widetext}
where $f_Q=\frac{df}{dQ}$. Besides, we can also take the variation of \eqref{1} with respect to the connection, yielding to 
\begin{equation}\label{8}
\nabla_{\mu}\nabla_{\gamma}\left( \sqrt{-g}f_Q {P^{\gamma}}_{\mu\nu}\right)=0\,.
\end{equation}
Here we are going to consider the standard Friedmann-Lema\^{\i}tre-Robertson-Walker (FLRW) line element, which is explicit written as
\begin{equation}
\label{9}
ds^2=-dt^2+a^2(t)\delta_{\mu\nu}dx^{\mu}dx^{\nu}\,,
\end{equation}
where $a(t)$ is the scale factor of the Universe. The previous line element enable us to write the trace of the nonmetricity tensor as
\begin{align*}
Q=6H^2\,.
\end{align*} 
Now, let us take the energy-momentum tensor for a perfect fluid, or
\begin{equation}
\label{10}
T_{\mu\nu}=(p+\rho)u_{\mu}u_{\nu}+pg_{\mu\nu}\,,
\end{equation}
where $p$ represents the pressure and $\rho$ represents the energy density. Therefore, by substituting \eqref{9}, and \eqref{10} in \eqref{7} one can find 
\begin{equation}
\label{11}
3H^2=\frac{1}{2f_Q}\left(-\rho+\frac{f}{2}\right)\,,
\end{equation}
\begin{equation}
\label{12}
\dot{H}+3H^2+\frac{\dot{f_Q}}{f_Q}H=\frac{1}{2f_Q}\left(p+\frac{f}{2}\right)\,,
\end{equation}
as the modified Friedmann equations for $f(Q)$ gravity. Here dot $(^.)$ represents one derivative with respect to time.

\section{Cosmographic Parameters}\label{sec4}

Modern cosmology is growing by a prominent number of observations. Therefore, the reconstruction of the Hubble diagram (i.e., the redshift- distance relation) is possible for higher redshift. The parametrization technique is a good method to study cosmological models. But, this type of procedure is entirely dependent on the models, and we check the viability by contrasting it against the observational data and putting limits on its model parameters. So, there are some unclear doubts about its characterizing parameters for the present-day values of the age of the Universe and the cosmological quantities. To overcome all these issues, one may adopt to cosmography. Cosmography is the study of scale factor by expanding it through the Taylor series with respect to the cosmic time. This type of expansion gives us distance-redshift relation and also independent of the solution of the motion equations of the cosmological models. To study cosmography, it is worthy of introducing the cosmographic parameters as follows:
\begin{equation}
\label{18}
H=\frac{1}{a}\frac{da}{dt},
\end{equation}
\begin{equation}
\label{19}
q=-\frac{1}{a}\frac{d^2a}{dt^2}H^{-2},
\end{equation}
\begin{equation}
\label{20}
j=\frac{1}{a}\frac{d^3a}{dt^3}H^{-3},
\end{equation}
\begin{equation}
\label{21}
s=\frac{1}{a}\frac{d^4a}{dt^4}H^{-4},
\end{equation}
\begin{equation}
\label{22}
l=\frac{1}{a}\frac{d^5a}{dt^5}H^{-5},
\end{equation}
where $H, q, j, s,$ and $l$ are represented the Hubble parameter, deceleration parameter, jerk parameter, snap parameter, and lerk parameter, respectively. These quantities are completely independent of model.\\
After some algebraic computation on Eqn \eqref{18}-\eqref{22}, one can derive the following relations:
\begin{equation}
\label{23}
\dot{H}=-H^2(1+q),
\end{equation}
\begin{equation}
\label{24}
\ddot{H}=H^3(j+3q+2),
\end{equation}
\begin{equation}
\label{25}
\dddot{H}=H^4[s-4j-3q(q+4)-6],
\end{equation}
\begin{equation}
\label{26}
H^{(iv)}=H^5[l-5s+10(q+2)j+30(q+2)q+24],
\end{equation}
where a dot $(^.)$ represents the one time derivative with respect to cosmic time $(t)$ and $H^{(iv)}=\frac{d^4H}{dt^4}$.\\
The degeneracy problem is one of the most common issues of the cosmological models. Cosmography is one of the best methods to deal with it. Furthermore, another advantage of cosmography is the luminosity distance can relate to the cosmographic parameters. In this concern, the direct measurement of luminosity distance can overcome the statistical error propagations, which has discussed in \cite{Cattoen/2008}. Therefore, the theoretical prediction can directly comparable to the observed data, without assuming a priori form of $H$ and $f(Q)$ \cite{will/2006}. The idea cosmography was extensively studied by S. Capozziello and his group in a modified $f(R)$ and $f(T)$ gravity. They also proposed a novel approach built on Pad\'e and Chebyshev polynomials to overcome standard cosmography limits based on Taylor expansion. Besides, they did a numerical analysis to constraints the functions of cosmographic sets using MCMC simulation \cite{capozziello/2019}. \\
The series expansion of the scale factor $a(t)$ up to its 5th order in terms of cosmographic set is
\begin{multline}
\label{27}
a(t)=a(t_0)\left[ H_0(t-t_0)-\frac{q_0}{2}H_0^2(t-t_0)^2 \right. \\ \left.
+\frac{j_0}{3!}H_0^3(t-t_0)^3+\frac{s_0}{4!}H_0^4(t-t_0)^4 \right. \\ \left.
+\frac{l_0}{5!}H_0^5(t-t_0)^5+\mathcal{O}[(t-t_0)^6]\right].
\end{multline}
By definition, the scale factor and redshift relation reads
\begin{equation}
\label{28}
\frac{a(t)}{a(t_0)}=\frac{1}{1+z},
\end{equation}
and the luminosity distance reads
\begin{equation}
\label{29}
d_L=\sqrt{\frac{\mathcal{L}}{4\pi\mathcal{F}}}=\frac{r_0}{a(t)},
\end{equation}
where $\mathcal{L}$ and $\mathcal{F}$ are the luminosity and flux, respectively. And,
\begin{equation}
\label{30}
r_0=\int_t^{t_0}\frac{d\eta}{a(\eta)},
\end{equation}
its physical meaning is the distance travels by a photon from a source at $r=r_0$ to the observer at $r=0$. Now, one can express the luminosity distance as series expansion of redshift $z$ with the cosmographic set and, also in terms of $f(z)$ and its derivatives, those are written in Appendix. Moreover, we can express $f(Q)=f(Q(z))=f(z)$ in terms of cosmograpgic set i.e., $f(z)=f(H(z),q(z),j(z),s(z),l(z))$. To do this,
we rewrite $Q$ in terms of redshift $z$ as
\begin{equation}
\label{31}
Q(z)=6H(z)^2,
\end{equation}
using definition of redshift in terms of cosmic time
\begin{equation}
\label{32}
\frac{d\log (1+z)}{dt}=-H(z).
\end{equation}
Therefore, we are able to calculate $Q$ and its derivatives with respect to $z$  and presented them in $z=0$. We ended up with the following results
\begin{equation}
\label{33}
Q_{0}=6 H_0,
\end{equation}
\begin{equation}
\label{34}
Q_{z0}=12 H_0 H_{z0},
\end{equation}
\begin{equation}
\label{35}
Q_{2z0}=12[H_{z0}^2+H_0H_{2z0}],
\end{equation}
\begin{equation}
\label{36}
Q_{3z0}=12[3H_{z0}H_{2z0}+H_0H_{3z0}],
\end{equation}
\begin{equation}
\label{37}
Q_{4z0}=12[3H_{2z0}^2+4H_0H_{3z0}+H_0H_{4z0}]
\end{equation}
Here, $Q_0=Q(z)|_{z=0}$, $Q_{z0}=\frac{dQ}{dz}|_{z=0}$, $Q_{2z0}=\frac{d^2Q}{dz^2}|_{z=0}$, etc. Similarly, $H_0=H(z)|_{z=0}$, $H_{z0}=\frac{dH}{dz}|_{z=0}$, $H_{2z0}=\frac{d^2H}{dz^2}|_{z=0}$, etc.

In order to express $Q$ and its derivatives in terms of cosmographic parameters, we have to evaluate the derivatives of $H(z)$ in terms of cosmographic parameters. To do so, we used \eqref{32} in \eqref{23}-\eqref{26} and got the following results
\begin{equation}
\label{38}
\frac{H_{z0}}{H_0}=1+q_0,
\end{equation}
\begin{equation}
\label{39}
\frac{H_{2z0}}{H_0}=j_0-q_0^2,
\end{equation}
\begin{equation}
\label{40}
\frac{H_{3z0}}{H_0}=-3j_0-4j_0q_0+q_0^2+3q_0^3-s_0,
\end{equation}
\begin{multline}
\label{41}
\frac{H_{4z0}}{H_0}=12j_0-4j_0^2+l_0+32j_0q_0-12q_0^2+25j_0q_0^2\\
-24q_0^3-15q_0^4+8s_0+7q_0s_0,
\end{multline}
Then using above equations, we are able to express $Q$ and its derivatives in terms of cosmographic set.
\section{$f(z)$ derivatives vs cosmography}\label{sec5}
As discussed above, the study of cosmological models by presuming an arbitrary form of $f(Q)$ and then solving the modified Friedmann equations creates doubt on its model parameters. So, in this section we try to express the derivatives of $f(z)$ in terms of the present values of the cosmographic parameters $(q_0,j_0,s_0,l_0)$. Doing this, gives us a hint about the functional form of $f(Q)$ which could be able to compit the observation.\\

\begin{figure*}[htbp]
	\centering
	\includegraphics[scale=0.6]{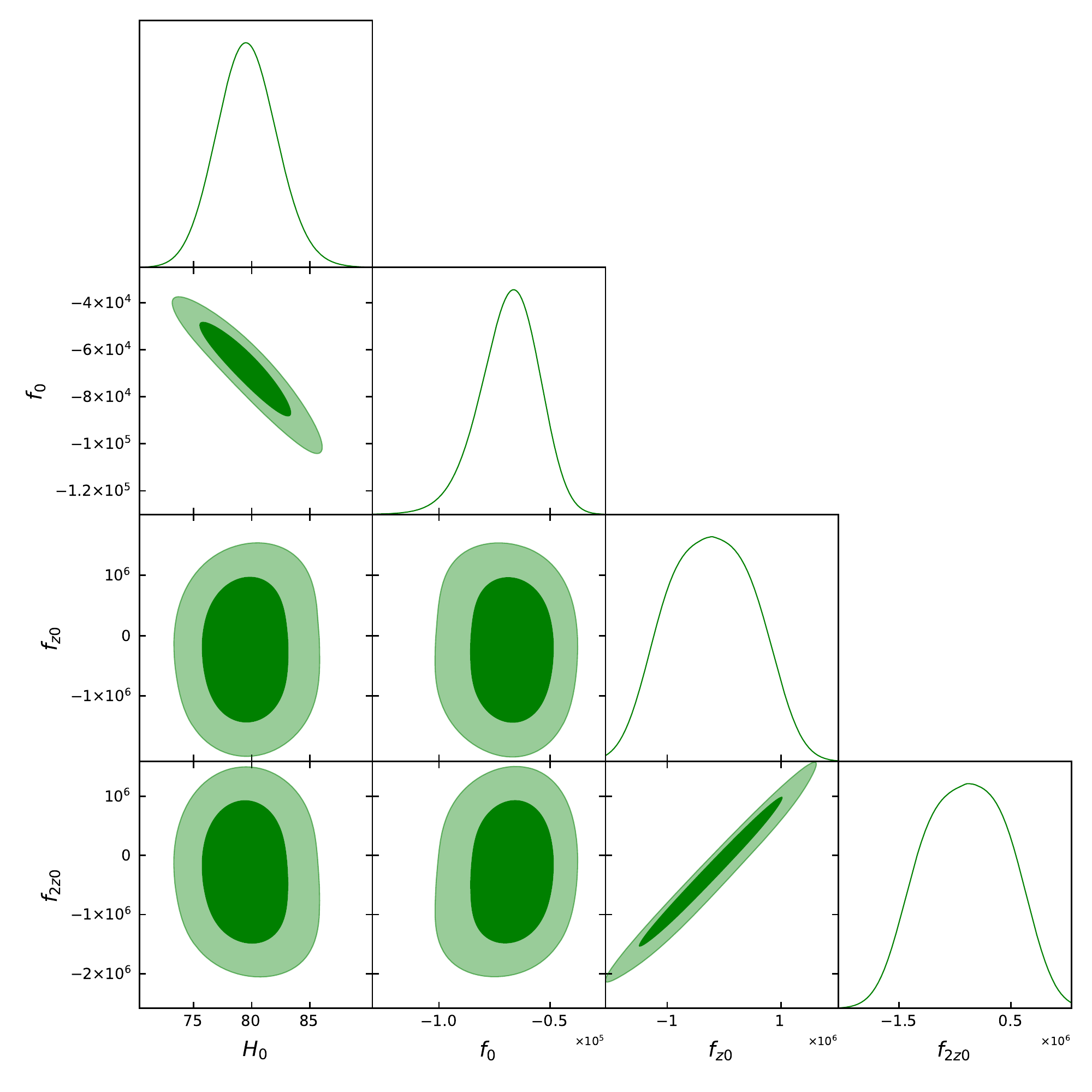}
	\caption{The marginalized constraints on the cosmographic parameters of M1 are shown by using the Pantheon SNe Ia sample. }
	\label{f1}
\end{figure*}

\begin{figure*}[htbp]
	\centering
	\includegraphics[scale=0.6]{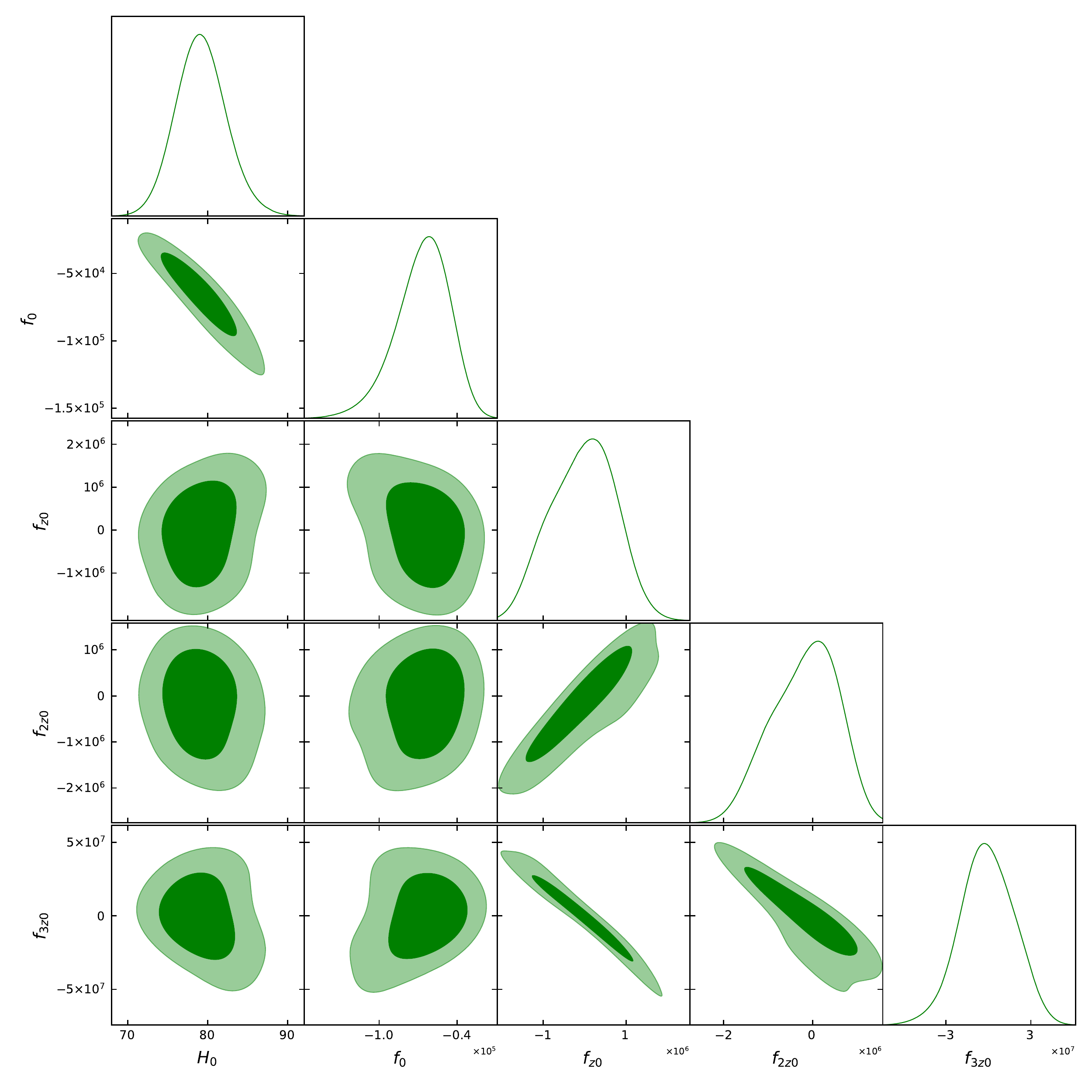}
	\caption{The marginalized constraints on the cosmographic parameters of M2 are shown by using the Pantheon SNe Ia sample. }
	\label{f2}
\end{figure*}

\begin{figure*}[htbp]
	\centering
	\includegraphics[scale=0.6]{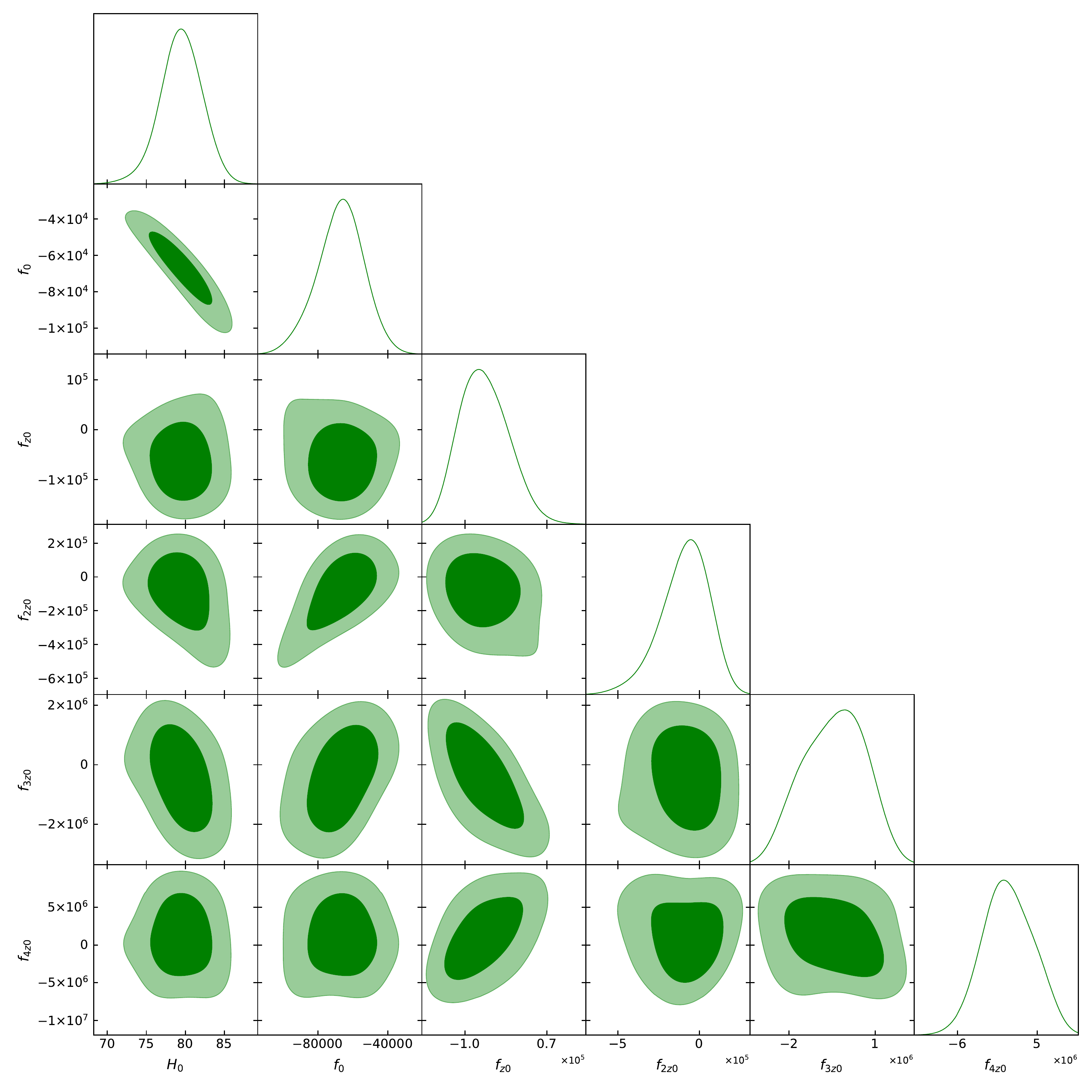}
	\caption{The marginalized constraints on the cosmographic parameters of M3 are shown by using the Pantheon SNe Ia sample. }
	\label{f3}
\end{figure*}

The modified motion Eqs. \eqref{11} and \eqref{12} can be rewritten as
\begin{equation}
\label{42}
H^2=\frac{1}{12 f'(Q)}[-Q\Omega_m+f(Q)],
\end{equation}
\begin{equation}
\label{43}
\dot{H}=\frac{1}{4f'(Q)}[Q\Omega_m-4H\dot{Q}f''(Q)],
\end{equation}
where $\Omega_m$ represents the dimensionless matter density parameter.\\
The $f(z)$ derivatives can be written as the functional dependence

\begin{align*}
f_z=f'(Q)Q_z,
\end{align*}
\begin{align*}
f_{2z}=f''(Q)Q_z^2+f'(Q)Q_{2z},
\end{align*}
\begin{align}
\label{44}
f_{3z}=f'''(Q)Q_z^3+3f''(Q)Q_zQ_{2z}+f'(Q)Q_{3z}.
\end{align}
and so on. Furthermore, following \cite{Lazkoz/2019,mandal/2020}, we know that $f(Q)=-Q$ mimic $\Lambda$CDM. Now, We are going to compare our results with the $\Lambda$CDM by fixing the bounds on $\Lambda$CDM
\begin{equation}
\label{45}
\Omega_{m0}=\frac{2}{3}(1+q_0),\hspace{0.5cm} f'(Q_0)=-1
\end{equation}
Using \eqref{45} in \eqref{42} we get
\begin{equation}
\label{46}
\frac{f_0}{6H_0^2}=\Omega_{m0}-2,
\end{equation}
\begin{equation}
\label{47}
\frac{f_{z0}}{6H_0^2}=-\frac{Q_{z0}}{6H_0^2},
\end{equation}
\begin{equation}
\label{48}
\frac{f_{2z0}}{6H_0^2}=-\frac{Q_{2z0}}{6H_0^2},
\end{equation}
and so on. Now, we can write the $f(z)$ and its derivatives as
\begin{equation}
\label{49}
\frac{f_0}{4H_0^2}=-2+q_0,
\end{equation}
\begin{equation}
\label{50}
\frac{f_{z0}}{12H_0^2}=-1-q_0,
\end{equation}
\begin{equation}
\label{51}
\frac{f_{2z0}}{12H_0^2}=-1-2q_0-j_0,
\end{equation}
\begin{equation}
\label{52}
\frac{f_{3z0}}{12H_0^2}=3q_0+q_0j_0-q_0^2+s_0,
\end{equation}
\begin{multline}
\label{53}
\frac{f_{4z0}}{12H_0^2}=-j_0^2-12q_0^4+19j_0q_0^2+16j_0q_0\\
-8q_0^2-12q_0^3+4s_0+l_0+7q_0s_0,
\end{multline}
Now, our aim is to put constraints on the vlues of $f_0$, $f_{z0}$, $f_{2z0}$, $f_{3z0}$, and $f_{4z0}$. In oder to do this, we have expressed the luminosity distance in terms cosmographic parameters as well as $f(z)$ and its derivatives for the present time in Appendix \ref{a}.

\section{Observational constraints}\label{sec6}

In this section, we deal with the luminosity distance $d_L$ to constraint $H_0, f_0, f_{z0}, f_{2z0},f_{3z0}$, and $f_{4z0}$,  with the observational data. For this, we have presented three statistical models with different maximum orders of parameters; this method, commonly accepted in the literature, corresponds to a hierarchy of parameters. Now, we are going to constraint the following models:
\begin{equation}
\mathbf{M1}:=\{H_0, \quad f_0, \quad f_{z0}, \quad f_{2z0} \},   \label{54}
\end{equation}
\begin{equation}
\mathbf{M2}:=\{H_0, \quad f_0,  \quad f_{z0}, \quad f_{2z0}, \quad f_{3z0} \},   \label{55}
\end{equation}
\begin{equation}
\mathbf{M3}:=\{H_0, \quad f_0, \quad f_{z0}, \quad f_{2z0}, \quad f_{3z0}, \quad f_{4z0} \}. \label{56}
\end{equation}
The motivation of doing such a hierarchical analysis of the cosmographic functions is that the extension of the sampled distributions by adding more parameters is optimistically expected. The resulting numerical effects on the measured quantities lead to a large distribution error, due to the higher orders of the Taylor's expansion. We are concerned in measuring these effects and resolving the cosmographic functions limitations. The numerical study is done by the MCMC analysis using SNe Ia data. As we know, SNe Ia is a powerful distance indicator to study the background evolution of the universe. In this study, to implement the cosmological constraints, we use the largest ``Pantheon'' SNe Ia  sample, which integrates SNe Ia data from the Pan-STARRS1, SNLS, SDSS, low-z and HST surveys and contains 1049 spectroscopically confirmed data points in the redshift range $z \in [0.01, 2.3]$ \cite{Scolnic}. 
%
%

To perform the standard Bayesian analysis, we employ a Markov chain Monte Carlo method to obtain the posterior distributions of cosmographic parameters. The best fits of the parameters are maximized by using the probability function
\begin{equation}\label{57}
\mathcal{L}\propto \exp(-\chi^2/2),
\end{equation}
where $\chi^2$ is the \textit{pseudo chi-squared function} \cite{hobson/2009}.
The marginalized constraining results are displayed in Figs.\ref{f1}-\ref{f3} and Tab.\ref{t1}. In Tab.\ref{t1}, the best fits are shown by the maximum likelihood function of the samples; the cited errors represent the 68\% confidence limits. From Figs. \ref{f1}-\ref{f3}, one can see marginalised posteriors lose Gaussianity when we apply additional parameters to Model M1. We conclude that considering Model M3 over Model M2 has the benefit of having more details on the cosmographic $f(Q)$ parameters without expanding the dispersion; however, Model M3 is less suitable for post-statistical treatment.

\begin{table*}[!t]
	\renewcommand\arraystretch{1.5}
	\caption{The marginalized constraining results on three cosmographic $f(Q)$ models M1, M2 and M3 are shown by using the Pantheon SNe Ia sample. We quote $1\,\sigma$ (68$\%$) errors for all the parameters here.
	}
	\begin{tabular} { l |c| c |c }
		\hline
		\hline

		Model              & M1      &M2      &M3        \\
		\hline
		$H_0$ & $79.5\pm2.5$     & $79.2\pm3.1      $    &$79.5\pm2.6$      \\
		\hline
		$f_0$ & $-0.68^{+0.14}_{-0.12}\times 10^5 $     & $-0.66^{+0.23}_{-0.17}\times 10^5 $    &$-0.67^{+0.14}_{-0.12}\times 10^5 $      \\
		\hline
		$f_{z0}$ & $(-0.22\pm0.73)\times 10^5 $     & $-0.04^{+0.86}_{-0.74}\times 10^6 $    &$-0.61^{+0.46}_{-0.56}\times 10^5 $      \\
		\hline
		$f_{2z0}$ & $(-0.30\pm0.74)\times 10^6 $     & $-0.17^{+0.87}_{-0.66}\times 10^6 $    &$-0.87^{+1.7}_{-1.2}\times 10^5 $      \\
		\hline
		$f_{3z0}$ & ---    & $(0.1\pm 1.9)\times10^7        $    &$-4^{+12}_{-11}\times 10^5  $      \\
		\hline
		$f_{4z0}$ & ---     & ---  &$(1\pm3)\times 10^6 $       \\

	    \hline
		\hline
	\end{tabular}
	\label{t1}
\end{table*}

\section{Discussions and conclusions}\label{sec7}

Cosmography provides a legitimate instrument for investigating cosmic expansion without a cosmological model. The constraints on the cosmographic parameters $(q_0, j_0, s_0, l_0)$ obtained by fitting to SNe Ia data. Also, these data completely support the cosmological principle. In certainty, any cosmological model should predict the cosmographic parameter values, which align with these values. Such a supposition makes it clear why the study of cosmography allows, as an interpretation of the cosmic speed observed, to verify its viability.

In this manuscript, we have dealt with the reconstruction of the correct form of $f(Q)$ function in $f(Q)$ gravity using the cosmographic parameters. We use the cosmographic parameters as a tool to derive $f(z)$ and its derivatives (called functions of cosmographic set fCS) in terms of cosmographic parameters. Also, we estimated the bounds on fCS using statistical analysis analysis with the 1048 data points from Pantheon SNe Ia sample, which includes Pan-STARRS1, SNLS, SDDS, low-z, and HST surveys data points.

Once, we did the expressions for fCS in terms of cosmographic parameters. Then we rewrite the expression of Luminosity distance in terms of fCS. Now, one can easily constrain $f(z)$ and its derivatives using numerical analysis. In our manuscript, we have adopted the MCMC statistical analysis and found the numerical bounds on fCS with the largest Pantheon SNe Ia sample, which are presented in Table \ref{t1}.

In this profile, we are able to constrain the fCS with the cosmographic values. Using the Pantheon data, we obtain a relatively high $H_0$ value, which is basically stable around $H_0=79.5$ km s$^{-1}$ Mpc$^{-1}$ in all three cosmographic models. Furthermore. we give primary constraints on the cosmographic parameters and find that only $f_0$ is nonzero beyond $2\sigma$ confidence level. Indeed, this implies that the former two terms in Taylor expansion of luminosity distance work dominantly.

\section*{Acknowledgments}   S.M. acknowledges Department of Science \& Technology (DST), Govt. of India, New Delhi, for awarding Junior Research Fellowship (File No. DST/INSPIRE Fellowship/2018/IF180676). PKS acknowledges CSIR, New Delhi, India for financial support to carry out the Research project [No.03(1454)/19/EMR-II Dt.02/08/2019]. 

\appendix
\section{Luminosity distance in terms of cosmographic parameters}
\label{a}
In this Appendix, we have written the expressions for the luminosity distance $d_L$ in terms of cosmographic parameters and $f(z)$ with its derivatives. The expression of $d_L$ reads
\begin{widetext}
\begin{multline}
\label{a1}
d_L(z)=\frac{1}{H_0}\left[z+\frac{1}{2}(1-q_0)z^2-\frac{1}{6}(1-q_0+j_0-3q_0^2)z^3+\frac{1}{24}(2+5j_0-2q_0+10j_0q_0-15q_0^2-15q_0^3+s_0)z^4 \right. \\ \left.
+\left(-\frac{1}{20}-\frac{9j_0}{40}+\frac{j_0^2}{12}-\frac{l_0}{120}
+\frac{q_0}{20}-\frac{11 j_0q_0}{12}+\frac{27q_0^2}{40}-\frac{7j_0q_0^2}{8}+\frac{11q_0^3}{8}+\frac{7q_0^4}{8}-\frac{11s_0}{120}-\frac{q_0s_0}{8}\right)z^5+\mathcal{O}(z^6)\right]
\end{multline}
The above equation can write in terms of $f(z)$ and its derivatives as
\begin{multline}
\label{a2}
d_L(z)=\frac{1}{H_0}\left[z-\frac{4H_0^2+f_0}{8H_0^2}z^2+\frac{1}{288 H_0^4}\left(9 f_0^2+168 f_0 H_0^2-4 f_{z0} H_0^2+4 f_{2z0} H_0^2+720 H_0^4\right)z^3 \right. \\ \left.
+\frac{1}{4608H_0^6}\left(-45 f_0^3+16H_0^4 (-846 f_0+23 f_{z0}-23 f_{2z0}+f_{3z0})-12H_0^2 f_0 (113 f_0-3 f_{z0}+3 f_{2z0})-44160H_0^6\right)z^4 \right. \\ \left.
+\frac{1}{92160 H_0^8}[279 f_0^4+16 H_0^4 \left(11268 f_0^2-417 f_0 f_{z0}+417 f_0 f_{2z0}-8 f_0 f_{3z0}+3 f_{z0}^2-6 f_{z0} f_{2z0}+3 f_{2z0}^2\right)\right. \\ \left.
+12 f_0^2 H_0^2 (967 f_0-26 f_{z0}+26 f_{2z0})+64 H_0^6 (19350 f_0-549 f_{z0}+549 f_{2z0}-23 f_{3z0}-f_{4z0})+3162624 H_0^8]z^5
\right]
\end{multline}
\end{widetext}



\begin{thebibliography}{90}
\bibitem{riess/1998} A. G. Riess, et al., \textit{Astron. J.}, \textbf{116}, 1009 (1998);  S. Perlmutter, et al., \textit{Astrophys. J.}, \textbf{517}, 565 (1999); S. Perlmutter, et al., \textit{Astrophys. J.}, \textbf{598}, 102 (2003);  D. N. Spergel,  et al.,  \textit{Astrophys. J. Suppl. Ser.}, \textbf{148}, 175 (2003); A. G. Riess, et al., \textit{Astrophys. J.}, \textbf{659}, 98 (2007);  D. N. Spergel,  et al.,  \textit{Astrophys. J. Suppl. Ser.}, \textbf{170}, 377 (2007).

\bibitem{carroll/1992}S. M. Carroll et al., \textit{Annu. Rev. Astron. Astrophys.}, \textbf{30}, 499 (1992); V. Sahni, A. Starobinski, \ijmpd, \textbf{9}, 373 (2000).

\bibitem{yang/2010}R. J. Yang, S. N. Zhang, \mnras,  \textbf{407}, 1835 (2010).

\bibitem{veltan/2014} H. E. S. Velten et al., \epjc, \textbf{74}, 3160 (2014).

\bibitem{joyce/2016} A. Joyce et al., \textit{Ann. Rev. Nucl. Part. Sci.}, \textbf{66}, 95 (2016).

\bibitem{baker/2019} T. Baker et al., (2019), arXiv:1908.03430.

\bibitem{peracaula/2019}J. S. Peracaula et al., \textit{Astrophys. J.}, \textbf{886}, L6 (2019); S. Mandal et al., \textit{Phys. Dark Universe}, \textbf{28}, 100551 (2020); Y. Yerramsetti, et al., \epjc, \textbf{79}, 1020 (2020); D. Wang, \textit{Phys. Dark Universe}, \textbf{28}, 100545 (2020); D. Wang et al., \epjc, \textbf{79}, 211 (2019); S. Arora et al., \cqg, 37, 205022 (2020); S. Bhattacharjee and P. K. Sahoo, \textit{Phys. Dark Universe}, \textbf{28}, 100537 (2010).

\bibitem{hehl/1995}F. W. Hehl et al., \textit{Phys. Rep.} \textbf{258}, 1 (1995); T. Ortin, \textit{Gravity and Strings} (Cambridge University Press, Cambridge, England, 2015).

\bibitem{aldrovandi/2014}R. Aldrovandi and J. Pereira, \textit{Teleparallel Gravity. An Introduction}, Fundamental Theories of Physics Vol. 173 (Springer, Dordrecht, 2014).

\bibitem{maluf/2013}J. W. Maluf, \textit{Ann. Phys.}, \textbf{525}, 339 (2013).

\bibitem{haghani/2012} Z. Haghani et al., \jcap, \textbf{10}, 061 (2012).

\bibitem{haghani/2013}Z. Haghani et al., \prd, \textbf{88}, 044024 (2013).

\bibitem{nester/1999}J. M. Nester and H.-J. Yo, \textit{Chin. J. Phys.} \textbf{37}, 113 (1999).

\bibitem{ferraro/2008}R. Ferraro and F. Fiorini, \prd, \textbf{78}, 124019 (2008); E. V. Linder, \prd, \textbf{81}, 127301 (2010); Y.-F. Cai et al., \textit{Rep. Prog. Phys.} \textbf{79}, 106901 (2016); C.-Q. Geng et al., \textit{Phys. Lett. B}, \textbf{704}, 384 (2011); L. Jarv and A. Toporensky, \prd, \textbf{93}, 024051 (2016); J. B. Jim´enez et al., \prd, \textbf{98}, 044048 (2018); A. Conroy and T. Koivisto, \epjc, \textbf{78}, 923 (2018).

\bibitem{jarv/2018} L. J\"arv et al., \prd, \textbf{97}, 124025 (2018).

\bibitem{hehl/1976} F. W. Hehl, et al., \textit{Rev. Mod. Phys.}, \textbf{48}, 393 (1976).

\bibitem{Lazkoz/2019}R. Lazkoz et al.,\prd, \textbf{100} 104027 (2019).

\bibitem{mandal/2020} S. Mandal, et al., \prd, \textbf{102}, 024057  (2020).

\bibitem{jianbo/2019} J. Lu et al., \epjc, \textbf{79}, 530 (2019).

\bibitem{mikhel/2018} M. R\"unkla and O. Vilson, \prd, 98, 084034 (2018).

\bibitem{viktor/2020}V. Gakis, et al., \prd, \textbf{101}, 064024 (2020).

\bibitem{harko/2018} T. Harko, et al., \prd, \textbf{98}, 084043 (2018).

\bibitem{olmo/2015} G. J. Olmo and D. Rubiera-Garcia, \textit{Phys. Lett. B}, \textbf{740}, 73 (2015).

\bibitem{einstein/1916}A. Einstein, Hamiltonsches Prinzip und allgemeine Relativit\"atstheorie, K\"oniglich Preu$\mathcal{B}$ische, Akademie der Wissenschaften (Berlin), Sitzungsberichte (1916), pp. 1111-1116.

\bibitem{Jimenez/2018} J. B. Jim\'enez, L. Heisenberg, T. Koivisto, \prd, 
\textbf{98}, 044048 (2018).

\bibitem{tomi/2018} T. Koivisto, \textit{Int. J. Geom. Methods Mod. Phys.}, \textbf{15}, 1840006 (2018).

\bibitem{Cattoen/2008}C. Cattoen, M. Visser, \prd, \textbf{78}, 063501 (2008); A. Aviles, et al., \prd, \textbf{87}, 044012 (2013); A. Aviles, et al., \prd, \textbf{86}, 123516 (2012); S. Capozziello, et al., \prd, \textbf{78}, 063504 (2008). 

\bibitem{will/2006}C. M. Will, \textit{Living Rev. Relativity}, \textbf{9}, 3 (2006).

\bibitem{capozziello/2019} S. Capozziello et al., \ijmpd, \textbf{28}, 1930016 (2019); S. Capozziello et al., \mnras, \textbf{476}, 3924 (2018).


\bibitem{Scolnic} D.~M.~Scolnic {\it et al.}, Astrophys.\ J.\  {\bf 859}, 101 (2018).

\bibitem{hobson/2009} \textit{Bayesian Methods in Cosmology}, edited by M. P. Hobson, A. H. Jaffe, A. R. Liddle, P. Mukherjee, and D.Parkison (Cambridge University Press, Cambridge, 2009).

\end{thebibliography}
\end{document}